\documentclass[showpacs,secnumarabic,amssymb,nobibnotes,aps,prl,11pt]{revtex4-2}
\usepackage{amsmath}
\usepackage{amssymb}
\usepackage{graphicx}
\usepackage{tabularx}
\usepackage{url,hyperref}
\usepackage[usenames,dvipsnames]{xcolor}
\hypersetup{colorlinks=true, linkcolor=BrickRed, urlcolor=blue!50!black, citecolor=blue!50!black}
\usepackage[capitalize]{cleveref}
\usepackage{cleveref}
\usepackage{natbib}
\AtBeginDocument{}
\crefrangelabelformat{equation}{(#3#1#4)$-$(#5#2#6)}

\begin{document}

\title{Coarsening in the Long-range Ising Model with Conserved Dynamics}

\author{Soumik Ghosh}
\author{Subir K. Das}
\email{das@jncasr.ac.in}
\affiliation{Theoretical Sciences Unit and School of Advanced Materials, Jawaharlal Nehru Centre for Advanced 
Scientific Research, Jakkur P.O., Bangalore 560064, India}
\date{\today}
\begin{abstract}
While the kinetics of domain growth, even for conserved order-parameter dynamics, is widely studied for short-range inter-particle interactions, systems having long-range interactions are receiving attention only recently. Here we present results, for such dynamics, from a Monte Carlo (MC) study of the two-dimensional long-range Ising model, with critical compositions of up and down spins. The order parameter in the MC simulations was conserved via the incorporation of the Kawasaki spin-exchange method. The simulation results for domain growth, following quenches of the homogeneous systems to temperatures below the critical values $T_c$, were analyzed via finite-size scaling and other advanced methods. The outcomes reveal that the growths follow power-laws, with the exponent having interesting dependence on the range of interaction. Quite interstingly, when the range is above a cut-off, the exponent, for any given range, changes from a larger value to a smaller one, during the evolution process. While the corresponding values at late times match with certain theoretical predictions for the conserved order-parameter dynamics, the ones at the early times appear surprisingly high, quite close to the theoretical values for the nonconserved case.  
\end{abstract}


\maketitle

\section{I.  Introduction}
When quenched inside a miscibility gap, from a high-temperature homogeneous phase, a system, comprising of multiple components, undergoes phase transition \cite{Fisher1967, Bray2002, Puri_b, Onuki2002, Cugliandolo2015, landaubook}. It tries to reach a new equilibrium, having regions of different components coexisting with each other, via the formation of small domains that are rich in like particles. Given the fundamental and technological significance, understanding of the growth of these domains received significant importance \cite{Bray2002, Puri_b, Onuki2002, landaubook, Lifshitz1961, Huse1986, Amar1988, Majumder2011_pre, Midya2017_prl, Laradji1996, Testard2011, Chauhan2024, Majumder2023, Mullic2017, Corberi2017,  Perrot1994, Lindt2021, Kiran1994, Bailey1993, Jiang2013, Nose1994} in material science and statistical physics communities. Usually, the related length scale $(\ell)$, i.e., the average size of the domains, grows with time in power-law fashion. For systems having only short-range interactions, the growth behavior is vastly studied and well understood. In such situations, for conserved order-parameter case, in the absence of hydrodynamics, mimicking phase separation, say, in solid mixtures, one expects \cite{Bray2002, Puri_b, landaubook, Onuki2002, Lifshitz1961}
\begin{equation}
    \ell(t)\sim t^{\alpha},~ \rm with ~\alpha=\frac{1}{3}.
    \label{growth_exp_ch4_eq}
\end{equation}
This is confirmed via computational studies \cite{Huse1986, Amar1988, Das2002, Majumder2011_pre} with spin models such as the nearest neighbor Ising model (NNIM) \cite{Majumder2011_pre} and the Potts model \cite{landaubook}. The expectation in Eq. \eqref{growth_exp_ch4_eq} appears valid for equal as well as unequal proportions of the components in mixtures.

The long-range counterpart \cite{Bray1990, Bray1993}, e.g., of the Ising model, in $d$ space dimensions, where the interaction falls off with distance $r$ as $r^{-(d+\sigma)}$, has become a topic of much recent interest \cite{Christiansen2019, Corberi2019, Christiansen2020, Agrawal2022, Muller2022, Ghosh2024}. In this case, for conserved order-parameter dynamics, a theory suggests \cite{Bray1993} 
\begin{equation}
    \alpha = \frac{1}{2+\sigma},~ \rm for ~\sigma<1.
    \label{growth_lrim_eq}
\end{equation}
 However, for $\sigma>1$, long-range interaction has been stated to have no role in deciding the domain growth exponent and thus \cite{Bray1993}, $\alpha = 1/3$, which is the same as the NNIM case. While this general fact may differ from our knowledge of
 boundary of interaction range, as well as understanding of universality, obtained from studies of equilibrium critical phenomena \cite{Fisher1972, Sak1973}, computer simulations showed consistency \cite{Muller2022, Ghosh2024}. It is worth mentioning here that a theoretical expectation for the nonconserved order-parameter dynamics is \cite{Bray1993, Bray1990} 
\begin{equation}
    \alpha = \frac{1}{1+\sigma},~ \rm for ~~\sigma<1,
\end{equation}
which also was found to be consistent with simulation studies \cite{Christiansen2020, Christiansen2019}.
In this case, $\alpha=1/2$, for $\sigma>1$, same as the nonconserved dynamics for the NNIM \cite{Bray2002}.

Here we revisit the case of long-range conserved dynamics. The objective, primarily, is to quantify the early-time behavior. 
For this purpose, we carry out Monte Carlo (MC) simulations \cite{landaubook}
%Analyses of our results, 
with the long-range Ising model (LRIM) \cite{Bray1993}, with equal proportion of mixture components. 
We analyze the obtained results via advanced methods, including a finite-size scaling (FSS) technique. The outcomes
reveal the following. For $\sigma$ somewhat less than unity, at early times we find the exponent $\alpha$ to be as high as even the theoretical expectation for the nonconserved dynamics. 
This is despite the fact that the imposed dynamics in our simulations is the conserved one. At late times, however, we find $\alpha \simeq 1/(2+\sigma)$, expectation recorded in Eq. \eqref{growth_lrim_eq} for the conserved dynamics. For $\sigma>1$, we find the growth to be consistent with $\alpha=1/3$, from rather early times. Our results on such crossover are surprising and, to the best of our knowledge, were never reported before.

The rest of the paper is organized as follows. Section II contains the descriptions of the model and the basic methods. Results, along with the discussions of the techniques of analysis, are presented in Section III. Finally, Section IV concludes the paper with a brief summary.

\section{II. Model and Methods}

We have studied the LRIM in $d=2$. The model has the Hamiltonian \cite{Christiansen2020, Ghosh2024, Bray1993, Bray1990}
\begin{equation}
    H=-\frac{1}{2}\sum_i\sum_{j\ne i}J(r)S_iS_j, ~ J(r)=\frac{1}{r^{d+\sigma}}.
\end{equation}
Here $J(r)$ is the interaction strength for a separation $r$ between two spins $S_i$ and $S_j$ at sites $i$ and $j$ that can take values $+1$ or $-1$, corresponding to, say, $A$ and $B$ types of particles in a $A+B$ binary mixture. For fixed $d$, $\sigma$ decides the range of interaction, for which we consider a wide variation.   

We start with random initial configurations, with $50:50$ up and down spins sitting on $2$D periodic $L\times L$ square lattices, mimicking a very high temperature ($T$) scenario. These configurations are instantaneously quenched to final temperatures $T_f=0.6T_c$, where $T_c$ represents a critical temperature \cite{Horita2017}. It should be noted that the value of $T_c$ depends upon $\sigma$, as well as on the system size $L$ \cite{Horita2017}.

We implement Kawasaki spin-exchange dynamics \cite{Bray2002} in our MC simulations \cite{landaubook, Frankelbook}. 
A trial move in this method is made of the following steps. First, a lattice site is randomly chosen. Spin (or a particle) at that site is then interchanged with that at a randomly chosen nearest neighbor site. This conserves the order parameter. Such a move is accepted according to the standard Metropolis criterion \cite{landaubook}. An MC step (MCS), the unit of our simulation time, is made of $L^2$ such moves. 

Given that our primary goal is to probe early time behavior, we have studied certain small system sizes, viz., $L=16,24$ and $32$. To identify the crossovers, we have also considered systems as large as $L=256$. Given the difficulty with the simulations of systems posessing long-range interactions, reaching adequately long time and getting good statistics is already difficult for the latter system size.
This is despite the fact that we have used Ewald summation \cite{Horita2017} and parallelized our code.  

The lengths from the evolution snapshots were calculated as the first moments of the domain-size distribution functions, there the size of a domain being the separation between two consecutive domain boundaries along any given Cartesian direction \cite{Majumder2011_pre}. For final temperatures that are reasonably above zero, there can be significant noise in the snapshots. This may introduce errors in the calculations of $\ell$. To avoid such inaccuracy we have eliminated this noise, via the application of a majority spin rule \cite{Majumder2011_pre}. 

All quantitative results are presented after averaging over data from runs with many different independent initial configurations. The numbers fall in the range between $175$ and $192$, for $L$ lying between $256$ and $16$.

\section{III. Results}

\begin{figure}
\centering
\includegraphics*[width=0.48\textwidth, height=0.35\textwidth]{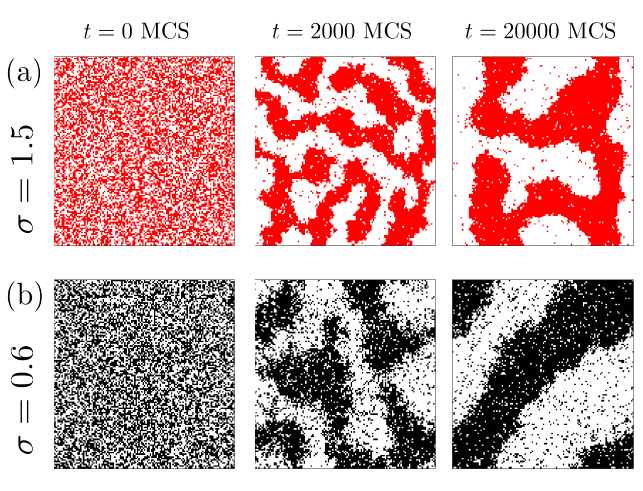}
\caption{
Evolution snapshots of the 2D systems, with $L=128$, when quenched to $0.6T_c$, from high temperature, are shown for (a) $\sigma = 1.5$ and (b) $\sigma = 0.6$. The locations of the particles are marked. For each $\sigma$ value, in addition to the initial configuration, two well-grown configurations are displayed. 
}
\label{snap_diff_sig_fig}
\end{figure}

\begin{figure}
\centering
\includegraphics*[width=0.45\textwidth, height=0.45\textwidth]{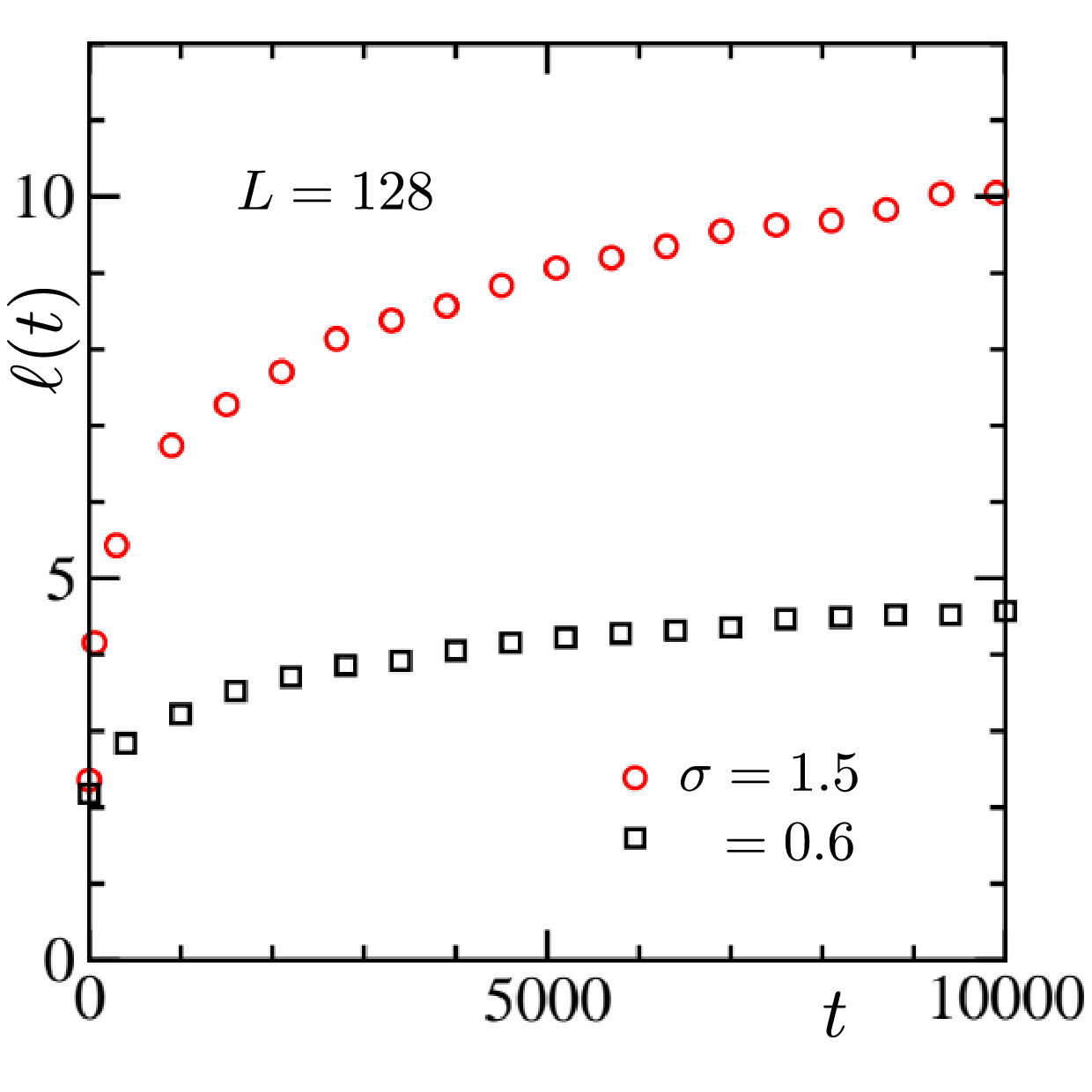}
\caption{
Domain length, $\ell(t)$, of systems with different $\sigma$ values, are plotted against time $t$. The system size is mentioned inside the frame. 
}
\label{dom_diff_sig_wn_fig}
\end{figure}

\begin{figure}
\centering
\includegraphics*[width=0.5\textwidth, height=0.41\textwidth]{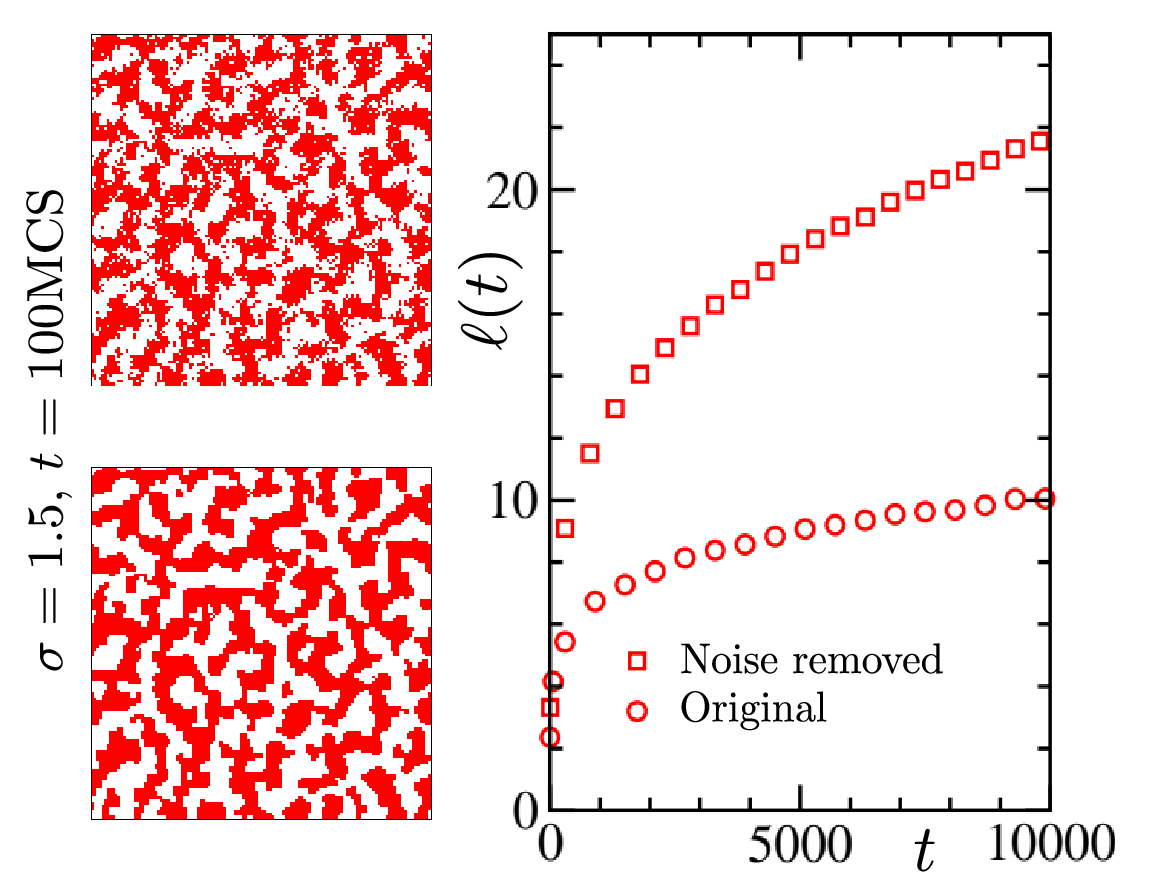}
\caption{
The upper left frame shows a snapshot taken during the evolution of a system with $\sigma=1.5$. The lower left frame contains the same snapshot after removing the noise. See the text for discussion and reference related to the removal of noise. The right frame shows a comparison between domain lengths calculated by using snapshots with and without noise.
These results are for $L=128$. 
}
\label{dom_wn_nr_sig_1.5_fig}
\end{figure}

In parts (a) and (b) of Fig. \ref{snap_diff_sig_fig}, we have shown evolution snapshots of systems with two different $\sigma$ values, viz., $\sigma = 1.5$ and $\sigma =0.6$, the former belonging to the short-range side of the interaction and the latter falling on the long-range side \cite{Bray1993}.
In each of the cases starting composition was a random arrangement of $A$ and $B$ particles. 
Clearly, for the longer range case the growth is faster. 
However, this cannot be concluded from Fig. \ref{dom_diff_sig_wn_fig} where we have shown plots of $\ell(t)$, versus $t$, for the same two values of $\sigma$. This unexpected quantitative feature is due to the fact that the snapshots for $\sigma=0.6$ contain significantly more noise than those for $\sigma=1.5$. 
In fact, for both the cases there exist much noise, the averages of which are time dependent till the corresponding length scales reach the values of equilibrium correlation lengths at the considered final temperatures. 
This fact on competing growth, between lengths related to noise and real coarsening, can lead to inappropriate conclusions on the rate of domain growth, that we are interested in, if the actual snapshots are used for the calculation of the latter.
In Fig. \ref{dom_wn_nr_sig_1.5_fig} we demonstrate how a noise removal exercise, mentioned earlier, picks up the actual domain morphology by discarding (noisy) fluctuations. See the two left frames and their descriptions in the caption. In the right frame, we plot $\ell(t)$, versus $t$, calculated by using snapshots with and without noise. 
Clearly, in the plot obtained by using the original snapshots, values of $\ell(t)$ are hugely underestimated! 
In the remaining part, thus, we have used the noise removed snapshots to calculate $\ell$. 
Our goal is to quantify the growths, for a large range of $\sigma$, the primary focus, as stated above, being on early times. 

\begin{figure}
\centering
\includegraphics[width=0.45\textwidth, height=0.45\textwidth]{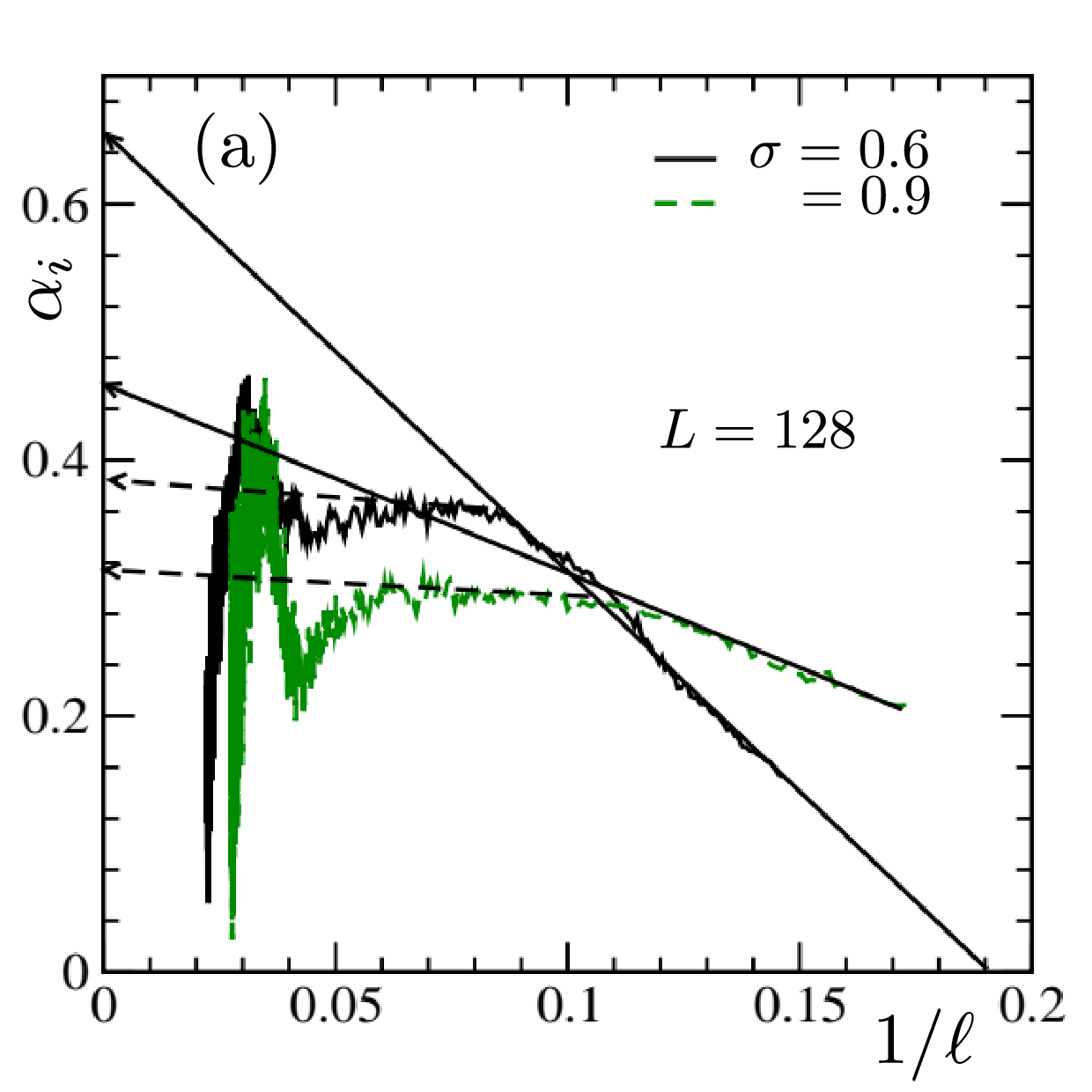}
\includegraphics[width=0.45\textwidth, height=0.445\textwidth]{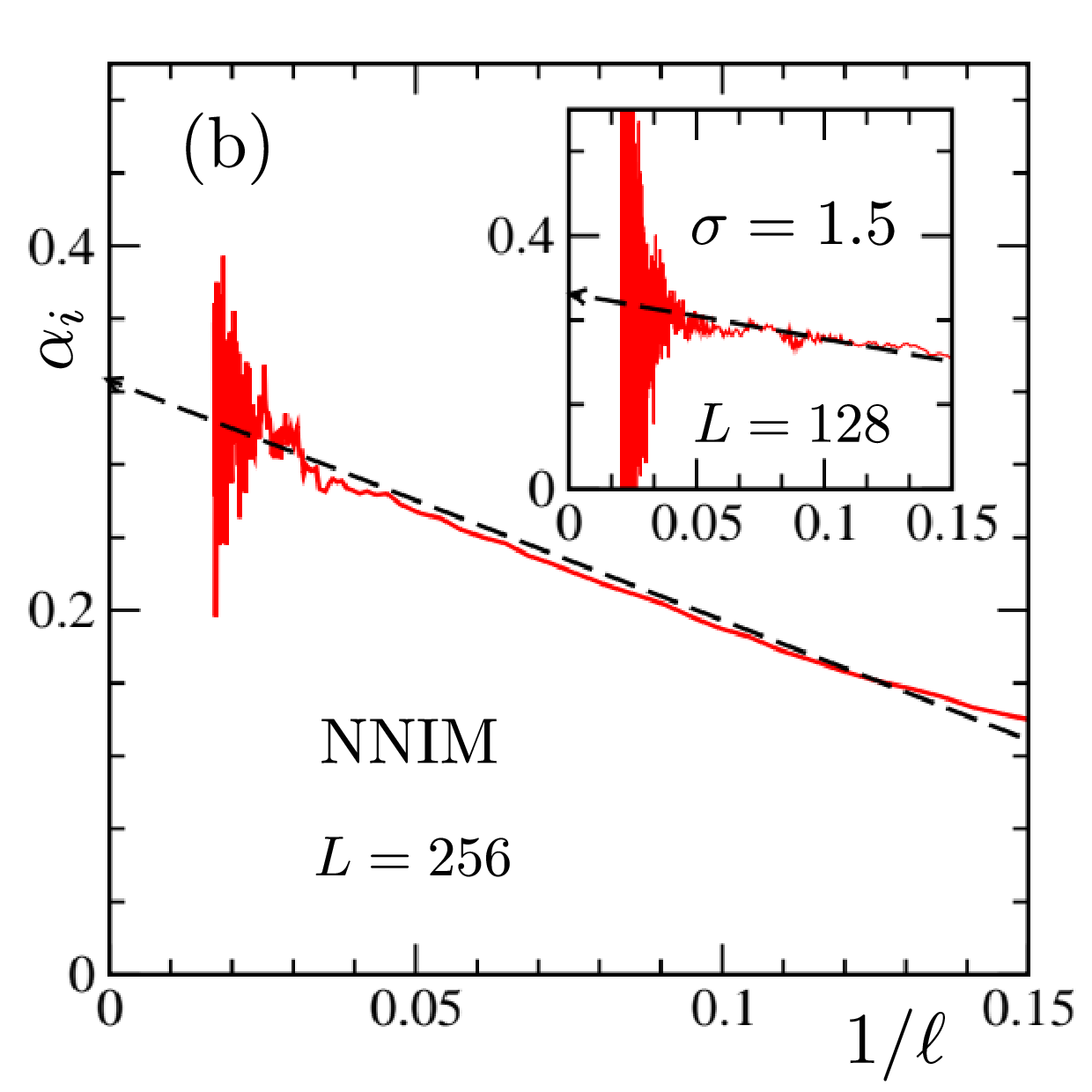}
\caption{
(a) Instantaneous exponent, $\alpha_i$, for different $\sigma$ values, viz., $\sigma=0.6$ and $0.9$, are plotted against $1/\ell$. (b) Same as (a), but here the results are for the nearest neighbor Ising model (NNIM). Inset shows the same plot for the LRIM with $\sigma=1.5$. Various arrow-headed lines are guides to the eyes. All the data are presented after running averaging. Data earlier than the presented ones suffer from noise removal exercise. See text for a discussion on the latter fact.
}
\label{inst_exp_diff_sig_nn_fig}
\end{figure}

We calculate the instantaneous exponent, $\alpha_i$, defined as \cite{Huse1986, Amar1988, Majumder2011_pre}
\begin{equation}
    \alpha_i = \frac{d(\ln\ell(t))}{d(\ln t)}.
    \label{alpha_i_ch4_eq}
\end{equation}
Value of $\alpha_i$, when $\ell \to \infty$, is the expectation for $\alpha$ in the long-time limit. In Fig. \ref{inst_exp_diff_sig_nn_fig}(a), we have plotted this quantity as a function of $1/\ell$, for $\sigma=0.6$ and $0.9$.  
It appears that there exist two distinct regimes, for each $\sigma$. Extrapolations of the trends exhibited by these two different regimes, to $\ell \to \infty$, i.e., $1/\ell \to 0$, lead to two drastically different values of $\alpha$. 
If we accept the late time trend, corresponding extrapolations are consistent with the theoretical predictions for conserved order-parameter dynamics \cite{Bray1993}, see Eq. \eqref{growth_lrim_eq}. On the other hand, extrapolations of the early time trends provide much higher values for $\alpha$!

In the main frame of Fig. \ref{inst_exp_diff_sig_nn_fig}(b), we have displayed a similar plot for the NNIM. In this case, the late time behavior is a continuation of rather early time trend \cite{Huse1986, Amar1988, Majumder2011_pre}. 
The same feature is true for the LRIM with $\sigma=1.5$ that already is on the short-range side of the interaction \cite{Bray1993}. 
See the inset for the latter plot. This hints towards the fact that the rates of growth, for the long-range variety, with $\sigma<1$, during early parts are different from the asymptotic values. Furthermore, careful observations suggest that the longevity of the early part keeps shortening before disappearing at a certain value of $\sigma$, possibly unity. In the following, we aim to confirm these early-time exponents via certain FSS analysis \cite{ Heermann1996, Fisher_barber, Majumder2011_pre}. 

%\begin{figure}
%\centering
%\includegraphics[width=0.47\textwidth, height=0.47\textwidth]{crossover_length.eps}
%\caption{
%Crossover lengths are plotted versus $\sigma$. The results are obtained with $L=128$.
%}
%\label{crossover_lengths_fig}
%\end{figure}

When a homogeneous system is quenched inside the coexistence regime, it takes a while to fall unstable and then reach a scaling regime of growth that may be considered free from any significant correction. If we denote this waiting time as $t_0$, and the corresponding length as $\ell_0$, $\ell(t)$ can be expressed as \cite{Amar1988, Majumder2011_pre}
    \begin{equation}
    \centering
        \ell(t) = \ell_0 + A{t^\prime}^\alpha,
        \label{t_0_l_0_eq}
    \end{equation}
with $t^\prime = t-t_0$. In simulations, due to the finiteness of the systems (and also due to certain freezing phenomenon \cite{Nalina2019}), domains can grow only up to a certain value, say, $\ell_{\rm max}$. In such situations, Eq. \eqref{t_0_l_0_eq} should be modified as, following a finite-size scaling \cite{Majumder2011_pre, Das2012, Heermann1996} ansatz,
\begin{equation}
    \centering
    \ell (t) - \ell_0 = Y(y) (\ell_{\rm max}-\ell_0).
    \label{scaling_ch4_eq}
\end{equation}
In Eq. \eqref{scaling_ch4_eq}, $Y(y)$ is a system-size independent (finite-size) scaling function and $y$ is a dimensionless scaling variable. A suitable choice for $y$ is \cite{Majumder2011_pre}
\begin{equation}
    \centering
    y = \frac{(\ell_{max}-\ell_0)^{\frac{1}{\alpha}}}{t^\prime}.
    \label{scaling_variable_ch4_eq}
\end{equation}
If we plot $Y(y)$ as a function of $y$, for the correct choices of $\ell_0$ and $\alpha$, data from different system sizes should collapse to form a master curve. In the limit $y \to \infty$, i.e., for the finite-size unaffected regime, a power-law behavior, 
\begin{equation}
    Y \sim y^{-\alpha},
    \label{scaling_fn_eq}
\end{equation}
should emerge, to comply with Eq. \eqref{t_0_l_0_eq}. 
However, if there exists a crossover between two different growth regimes, data collapse cannot be expected. Thus, to probe the early time behavior, study should be restricted to smaller system sizes. For this reason, for the FSS analysis, to identify early time growth exponent, we choose $L=16, 24$ and $32$, for which the finite-size effects enter before the crossover, for the considered $\sigma$ value, viz., $\sigma=0.6$, that we use to demonstrate the exercise.

\begin{figure}
\centering
\includegraphics[width=0.45\textwidth, height=0.45\textwidth]{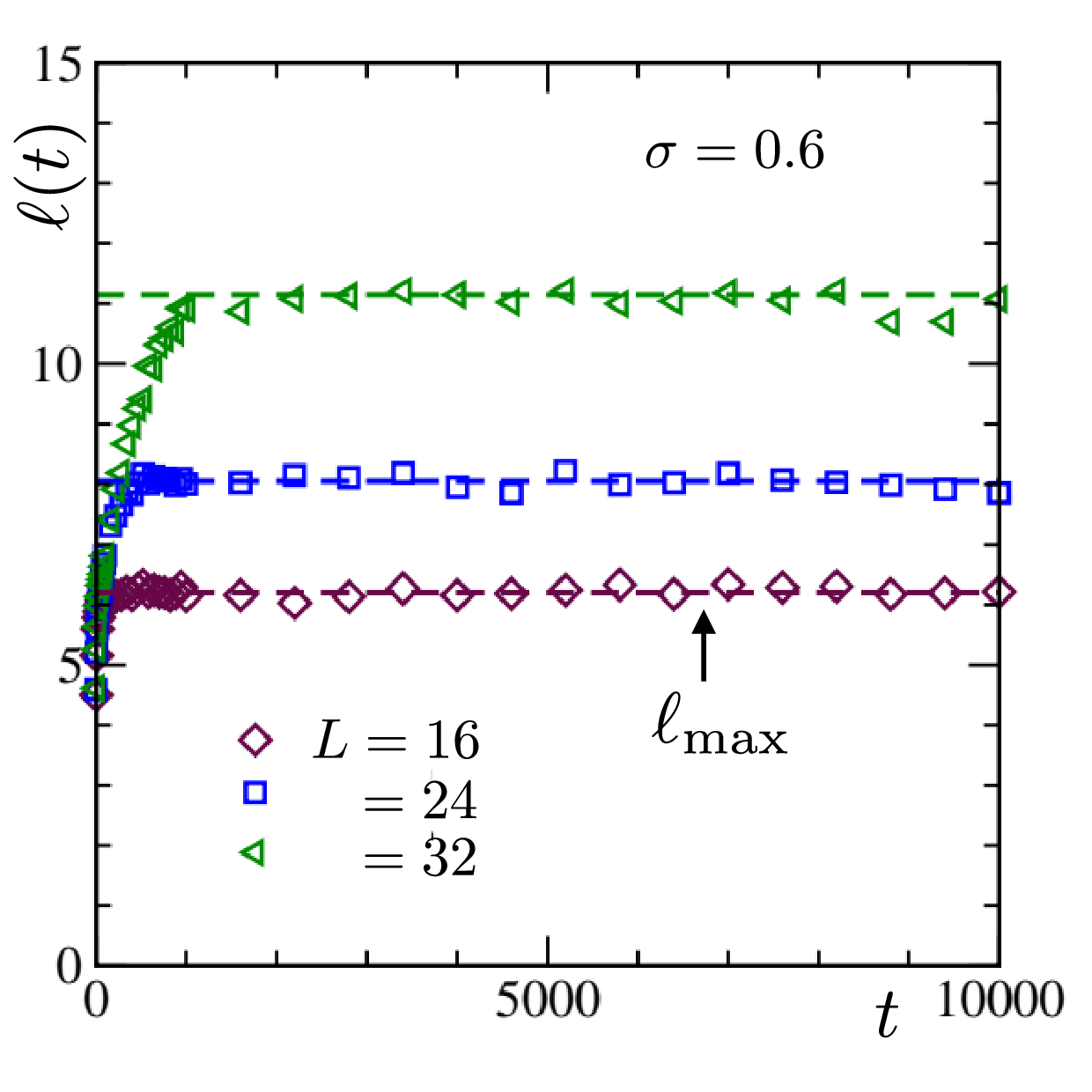}
\caption{
(a) Plots of $\ell(t)$, versus $t$, for $\sigma=0.6$, from three different system sizes. The dashed horizontal lines are our estimates for $\ell_{\rm max}$. 
See text for the definition of the latter.
}
\label{domain_diff_L_sig_0.6_fig}
\end{figure}
 
\begin{figure}
\includegraphics[width=0.45\textwidth, height=0.45\textwidth]{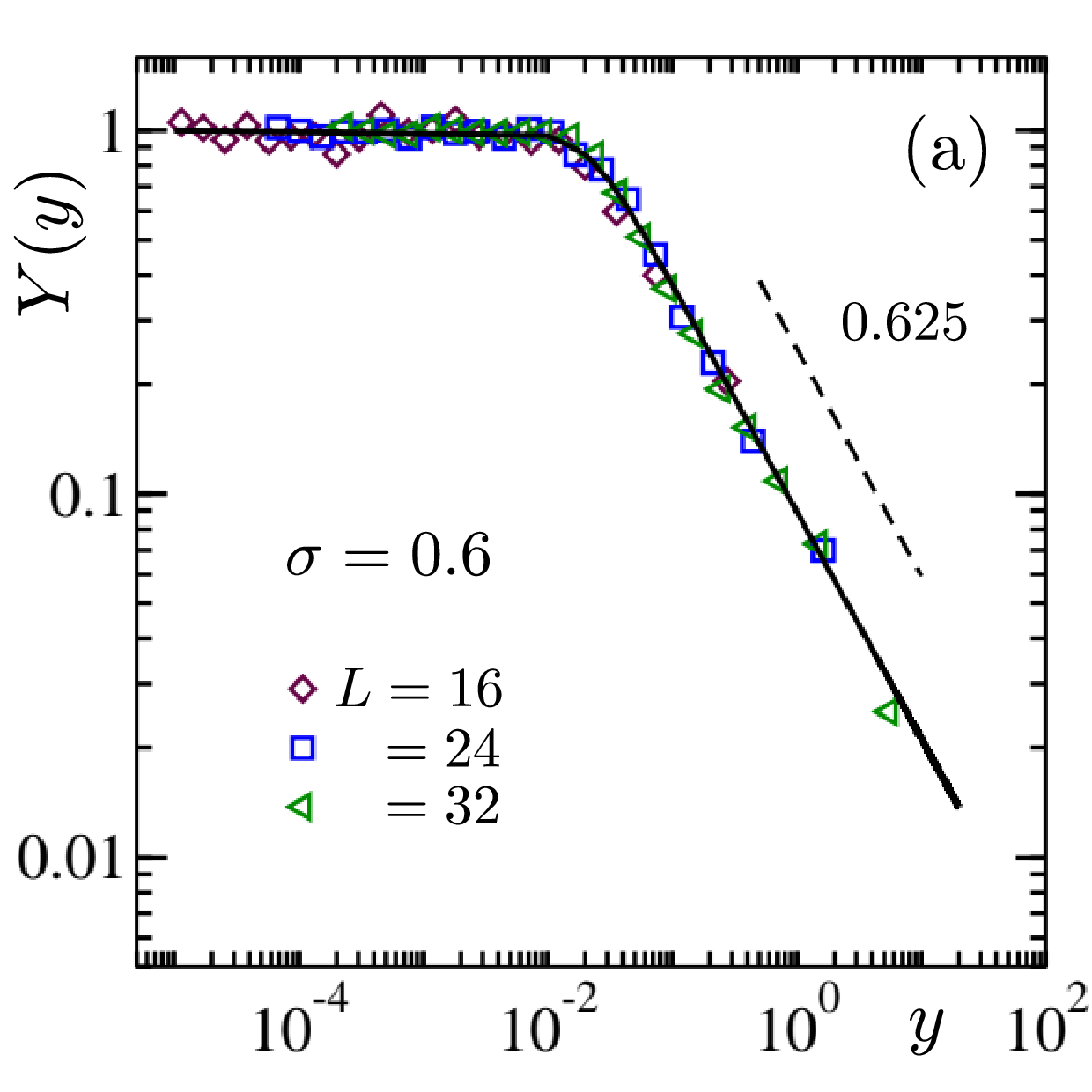}
\includegraphics[width=0.45\textwidth, height=0.445\textwidth]{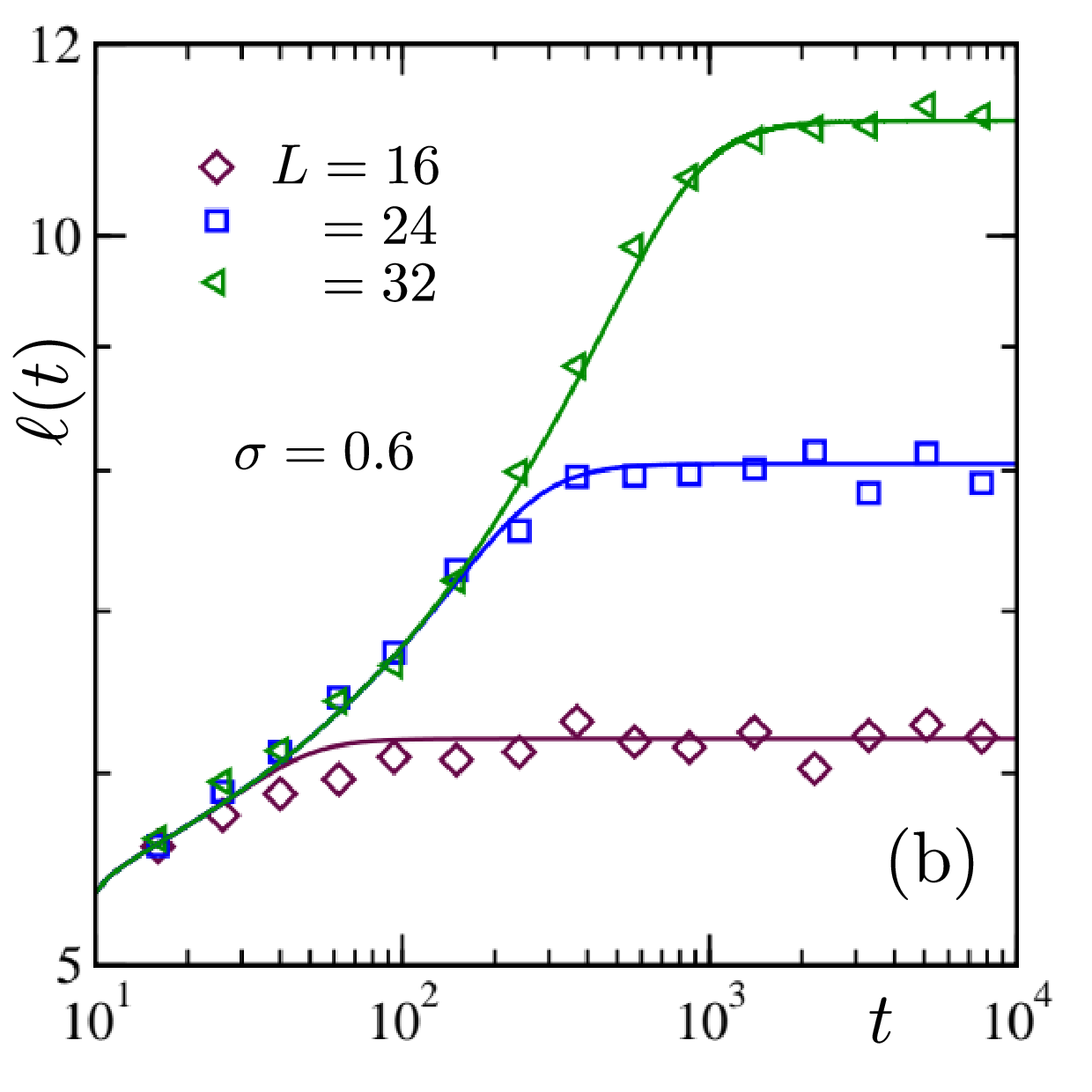}
\caption{
(a) Double-log plots of the finite-size scaling function, $Y(y)$, against the scaling variable $y$, for  $\sigma = 0.6$. Data for a few different system sizes, viz., $L = 16, 24$, and $32$, are included. The dashed line represents Eq. \eqref{scaling_fn_eq} with $\alpha=0.625$. The solid line is a fit of the simulation data to Eq. \eqref{analytic_scaling_fn_eq}. See text for the best fit values of relevant parameters. 
(b) Eq. \eqref{finite_domain_fn_eq} (see the solid lines) is compared with direct growth data for $L=16,24$ and $32$. For the purpose of clear visualization, of comparison between simulation data and the analytical lines, we have thinned down the data sets in both parts (a) and (b).
}
\label{finite_scaling_sig_0.6_fig}
\end{figure}

In Fig. \ref{domain_diff_L_sig_0.6_fig}, for $\sigma=0.6$, we have shown $\ell(t)$, as a function of $t$, for the above three system sizes. It is clear from the plots that for different system sizes finite-size effects appear at different times. 
In Fig. \ref{finite_scaling_sig_0.6_fig}, we have shown the finite-size scaling exercise, on a log-log scale, using the same set of system sizes.
A very good collapse of data is achieved for the choices $\alpha  \simeq 0.625$ and $\ell_0 = 5.35$. 
This value of $\alpha$ is the theoretically predicted exponent for the nonconserved dynamics with long-range interaction \cite{Bray1993, Christiansen2020}, viz., $\alpha = 1/(1+\sigma)$! 
For large $y$, the collapsed data are consistent with Eq. \eqref{scaling_fn_eq}, for $\alpha=0.625$, an expectation for the validity of the FSS analysis.

For $\sigma=0.6$, the observed value of $\alpha$, from the FSS, should be compared with $\simeq 0.65$, the number that emerges from the extrapolation of early time trend in Fig. \ref{inst_exp_diff_sig_nn_fig}(a). The agreement is reasonably close. Here it should be noted that for the NNIM the linear behavior of $\alpha_i$ was shown to have a connection with $\alpha$ and $\ell_0$ as \cite{Amar1988, Majumder2011_pre}
\begin{equation}
    \alpha_i = \alpha \left[ 1 - \frac{\ell_0}{\ell}\right].
    \label{form_alpha_i_eq}
\end{equation}
By considering the FSS numbers, viz., $\alpha=0.625$ and $\ell_0=5.35$, we obtain $-\alpha\ell_0$, the slope of $\alpha_i$ vs $1/\ell$ plot, to be $\simeq -3.34$. This is quite close to $-3.42$, the measured slope from Fig. \ref{inst_exp_diff_sig_nn_fig}(a), deviating from each other by less than even $3\%$. 

We have used an analytical form for the scaling function $Y(y)$, given in Ref. \cite{Das2021}:
\begin{equation}
    Y(y) = Y_0\left(b+\dfrac{y^{\theta}}{\alpha}\right)^{-\alpha/\theta}, 
    \label{analytic_scaling_fn_eq}
\end{equation}
where $Y_0$, $b$, and $\theta$ are positive constants. Here the value of $\theta$ should determine the finite-size universality class. 
While obtained for a somewhat different purpose, usefulness of this function for phase transitions in finite systems was recently demonstrated in Refs. \cite{Das2024, Paul2024}.
This function describes the scaled data in Fig. \ref{finite_scaling_sig_0.6_fig}(a), with $\theta \simeq 2.6$, quite nicely, while $\alpha=0.625$.

From Eq. \eqref{analytic_scaling_fn_eq}, via a back transformation, one can write \cite{Paul2024}
\begin{equation}
    \ell(t)=\ell_0+Y_0(\ell_{\rm max}-\ell_0)\left[ b + \frac{(\ell_{\rm max}-\ell_0)^{\theta/\alpha}}{\alpha(t-t_0)^{\theta}} \right]^{-\alpha/\theta}.
    \label{finite_domain_fn_eq}
\end{equation}
The analytical form in Eq.\eqref{finite_domain_fn_eq} can be used to describe direct data for growth in finite systems of arbitrary sizes. This usefulness is demonstrated in Fig. \ref{finite_scaling_sig_0.6_fig}(b).
 
\begin{figure}
\centering
\includegraphics[width=0.48\textwidth, height=0.4\textwidth]{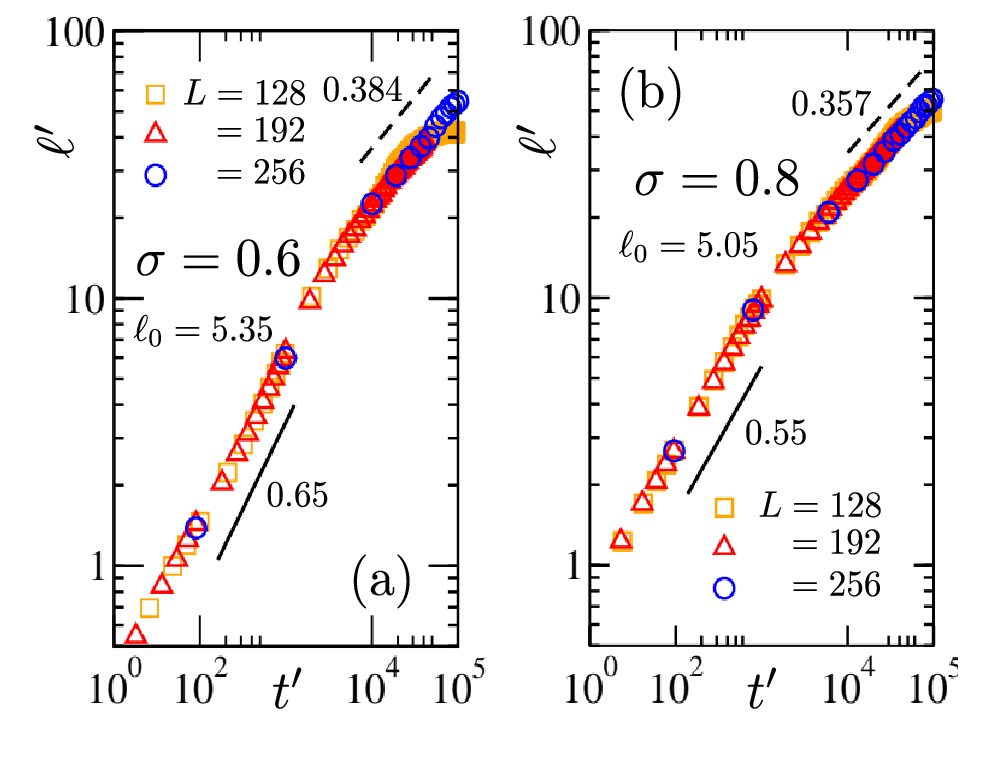}
\caption{
Plots of $\ell^\prime(t)~(=\ell-\ell_0)$, as a function of $t^\prime$ (=$t-t_0$), on a log-log scale, for $(a)$ $\sigma = 0.6$ and $(b)$ $\sigma = 0.8$. In each of the cases data from three different system sizes have been included. This is to show that the post-crossover bending is not due to finite-size effects. Solid lines are for the initial behavior of the growth and the dashed lines denote the later time, $\alpha=1/(2+\sigma)$, growth.  
}
\label{exp_cross_diff_sig_fig}
\end{figure}

Covering a wide range of $\sigma$, we intent to draw a complete picture, depicting two different growth regimes, before and after the crossover. For this purpose, we need bigger systems. In Fig. \ref{exp_cross_diff_sig_fig}(a), for $\sigma=0.6$, we have plotted $\ell^\prime~(= \ell(t)-\ell_0$) as a function of $t'$,  for relatively larger system sizes: $L=128, 192$ and $256$. As expected, on a log-log scale, we observe two distinct linear-looking regimes with a crossover, suggesting that the exponent changes from quite a high value to a lower one. 
In part (b) of this figure, we have presented similar plots, but here for $\sigma=0.8$, and observed a similar crossover.

\begin{figure}
\centering
\includegraphics[width=0.5\textwidth, height=0.25\textwidth]{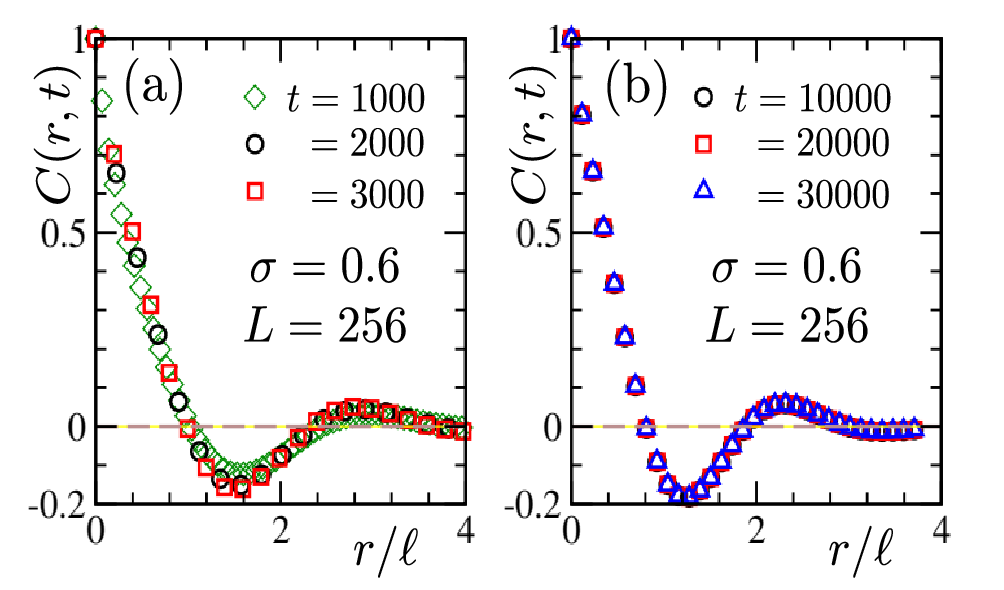}
\caption{
Scaling plot of $C(r,t)$, for $\sigma=0.6$ and $L=256$, using data from (a) early and (b) late times.
}
\label{corfn_early_late_fig}
\end{figure}

From these exercises, we conclude that for $\sigma<1$ initially one observes very rapid growth, which crosses over, at late times, to $1/(2+\sigma)$. 
To check if structural scaling exists in both the regimes, in Fig. \ref{corfn_early_late_fig} we show data collapse for 
\begin{equation}
C(r,t) = \langle S_{i}(t)S_{j}(t)\rangle - \langle S_{i}(t)\rangle\langle S_{j}(t)\rangle ,
\label{corfn_eq}
\end{equation}
the two-point equal-time correlation function \cite{Bray2002} for separation $r$ between lattice sites $i$ and $j$, when $\sigma=0.6$. Part (a) is for early period and part (b) contains data from late times. Nice collapse can be appreciated throughout. Here it should be noted that the observed scaling of the form \cite{Bray2002}
\begin{equation}
    C(r,t) \equiv C(r/\ell(t))
    \label{corfn_scaling_eq}
\end{equation}
implies self-similar growth. Furthermore, the validity of the scaling confirms that the comparison of $\ell$ at different times is meaningful.

%\begin{figure}
%\centering
%\includegraphics[width=0.48\textwidth, height=0.4\textwidth]{finite_scaling_sig_gt_1.eps}
%\caption{
%Double-log plots of scaling functions $Y(y)$, against the scaling variable $y$, for (a) $\sigma = 1.1$ and (b) $\sigma = 1.5$. Data for 2 different system sizes, viz., $L = 32$ and $64$, are used in both the cases. The solid lines are power-laws with exponent $-1/3$. The dashed lines are fits to the Eq. \ref{analytic_scaling_fn_eq}.
% }
%label{finite_scaling_gt_1_fig}
%\end{figure}

\begin{figure}
\centering
\includegraphics*[width=0.48\textwidth, height=0.4\textwidth]{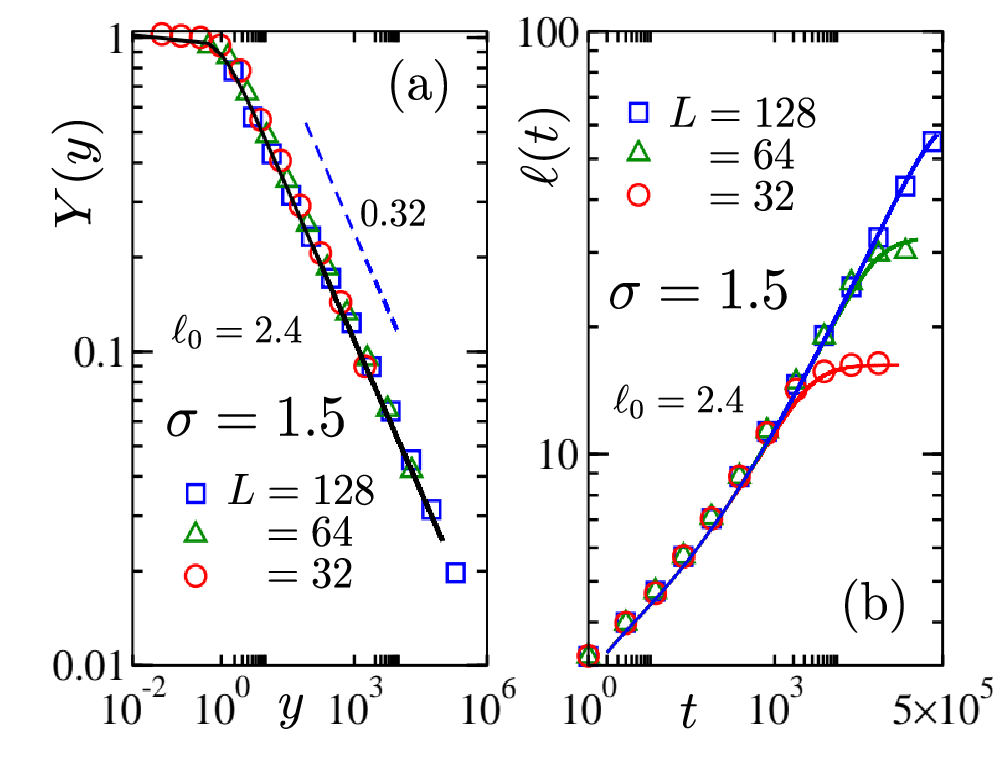}
\caption{
(a) Finite-size scaling exercise for $\sigma=1.5$, using $\ell(t)$ data from different system sizes. The dashed line represents Eq. \eqref{scaling_fn_eq} with $\alpha=0.32$ and the solid line is a fit of the simulation data to the analytical form in Eq. \eqref{analytic_scaling_fn_eq}. 
(b) Same as Fig. \ref{finite_scaling_sig_0.6_fig}(b) but here the demonstration is for $\sigma=1.5$. The considered values of $\ell_{\rm max}$ for $L=32, 64$ and $128$ are $16, 32$ and $64$, respectively. Like in Fig. \ref{finite_scaling_sig_0.6_fig} here also we have thinned down the data sets, for visual clearty. 
}
\label{finite_scaling_sig_1.5_fig}
\end{figure}

\begin{figure}
\centering
\includegraphics*[width=0.45\textwidth, height=0.45\textwidth]{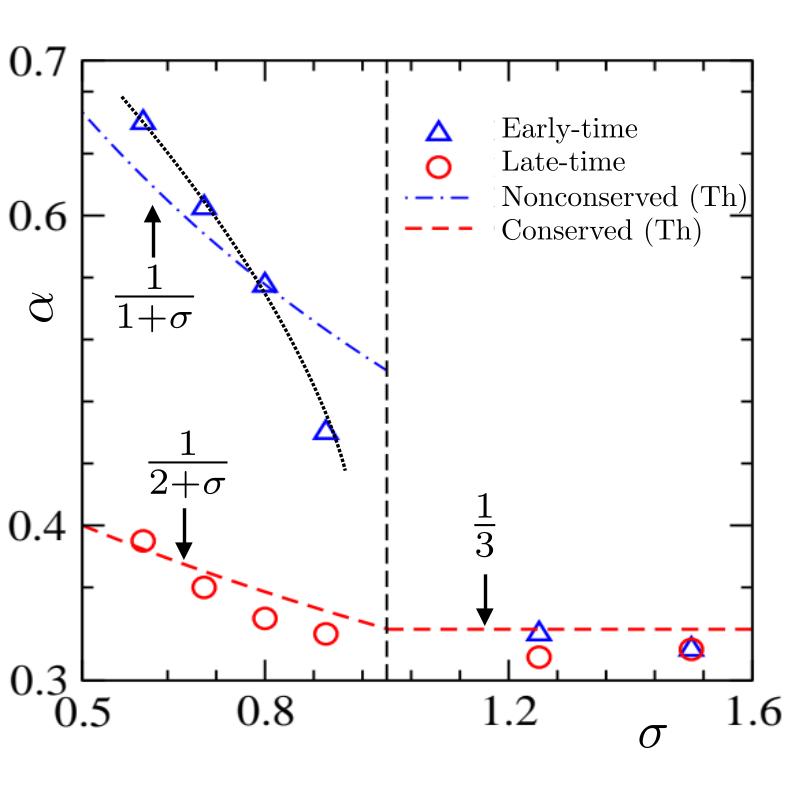}
\caption{
Both early-time and late-time values of the growth exponent $\alpha$ are plotted against $\sigma$. Simulation results are shown with symbols, whereas the relevant theoretical predictions (Th) are shown by dashed and dashed-dotted lines. The continuous line is a guide to the eye. The values of the late time exponents for $\sigma<1$ are taken from Ref. \cite{Ghosh2024}. For $\sigma>1$, it was difficult to choose regions of data sets to estimate early and late time values of $\alpha$. The overlapping numbers provide testimony to this fact.
}
\label{alpha_vs_sig_diff_reg_fig}
\end{figure}

Next, we visit the finite-size scaling again, this time for $\sigma>1$. Double-log plots of scaling function $Y(y)$, obtained by using different system sizes, are plotted in Fig. \ref{finite_scaling_sig_1.5_fig}(a), against the scaling variable $y$, for $\sigma = 1.5$. 
An excellent collapse is realized when $\alpha=0.32$, for small as well as large system sizes. 
The dashed line is a power-law with exponent $-0.32$. This is consistent with the conclusion from the exercise in the inset of Fig. \ref{inst_exp_diff_sig_nn_fig}(b). 
The solid line in Fig. \ref{finite_scaling_sig_1.5_fig}(a) represents a fit to the Eq. \eqref{analytic_scaling_fn_eq}. 
Given that here we have used larger systems, it is, thus, clear that the domain growth exponent $\alpha$ remains constant throughout with a value $\alpha \simeq 1/3$ for $\sigma > 1$ \cite{Bray1993}. 
Here also, in part (b), we demonstrate, how the direct data, for different system sizes, can be described by Eq. \eqref{finite_domain_fn_eq}.

Recalling the objective of drawing a comprehensive picture, we return to $\sigma<1$. 
Given the fact that as $\sigma$ approaches unity, the crossover occurs earlier and at a smaller length scale, it becomes a necessity to restrict simulations with even smaller systems for a FSS. In that case, however, FSS will suffer from significant corrections to scaling. This can already be appreciated from Fig. \ref{finite_scaling_sig_0.6_fig}(b). 
Clearly there exists a deviation between the analytical function [see Eq. \eqref{finite_domain_fn_eq}] and the simulation data for the smallest system size, whereas the agreement is nearly perfect for the larger systems. In such a situation, appreciating the fact that the FSS and the instantaneous exponent approaches provide values of $\alpha$ that are close to each other, we stick to the estimates from the latter method. 

In Fig. \ref{alpha_vs_sig_diff_reg_fig}, we provide a detailed picture of growth covering both long and short-range regimes of $\sigma$. It appears, as $\sigma \to 1$, the early time exponent tends to merge with the late time values at $\alpha \simeq 1/3$. For $\sigma$ reasonably small, interestingly, these values of $\alpha$ are quite consistent with the corresponding numbers for the nonconserved order-parameter dynamics.

\section{IV. Conclusion}

Although the problem of domain coarsening in different model systems, following quenches to state points inside the coexistence curve, received much importance, studies with long-range interactions have been carried out relatively rarely. Nevertheless, there exist theoretical predictions \cite{Bray1990, Bray1993} for the latter variety, though only a handful of studies considered confirming these predictions via simulations \cite{Christiansen2019, Corberi2019, Christiansen2020, Agrawal2022, Muller2022, Ghosh2024}. 
%For a system where interaction between two spins decay as a function of the radial distance $r$ as $1/r^{d+\sigma}$, $d$ being the space dimension, the growth exponent $\alpha$ is a function of $\sigma$ given $\sigma<1$. Whereas, for $\sigma>1$, the coarsening dynamics is the same as it can be seen in systems with short-range interactions, e.g. nearest-neighbor interaction. 
%In one of the previous chapters, we have shown the estimated values for the growth exponents from our simulations. In the thermodynamic limit, the trend matches the theoretical predictions \cite{Bray1993}. 
%A previous simulation study on $1$D long-range Ising model claimed \cite{Corberi2019} that for the conserved order parameter dynamics  $\alpha=1/(1+\sigma)$, for $\sigma<1$, despite the fact that the RG theory prediction for such system is $\alpha = 1/(2+\sigma)$. As the value of $\alpha$ is independent of the dimensionality, one may expect the same conclusion for $2$D as well.

In this work, we have carefully studied the domain coarsening in $2$D long-range Ising model, via computer simulations.
We have performed Monte Carlo simulations with the Kawasaki spin-exchange mechanism. Systems prepared at very high temperature were quenched to final temperatures $T=0.6T_c$, for various different values of $\sigma$, that dictate the range of interaction. 
From the analyses of the obtained simulation results, using finite-size scaling and other advancd techniques, we have come to the conclusion that for $\sigma<1$, there exist two distinct regimes of growth. 
Initially, domains grow much faster with exponents being close to $1/(1+\sigma)$, a prediction for nonconserved dynamics. Then, a ``crossover'' takes place, and $\alpha$ picks up the theoretically predicted value, $1/(2+\sigma)$, for the conserved order parameter. The early time exponent, however, tends to $1/3$ as $\sigma$ increases. 
For $\sigma>1$, from reasonably early times, the growth exponent picks up the theoretical number \cite{Bray1993} $1/3$, irrespective of the value of $\sigma$.

It may be recalled that we have removed the noise in the snapshots before calculating the average sizes of domains. Given that at early times the noise is relatively less, it may be a valid question to ask, whether the exceptionally fast growths at early times, for small values of $\sigma$, are due to the removal of the noise. Note that when the domains are small in size, this exercise may lead to artificial merger of these, leading to inappropriate conclusions. 
Keeping this in mind, we have analyzed the data for early periods without removing the noise as well. Our conclusions remain essentially unchanged. E.g., for $\sigma=0.6$, we get pre-crossover exponent to be $\simeq 0.6$. On the other hand, if the noise is not removed, post-crossover exponent for this $\sigma$ value appears much smaller than the expectation!

%It also seems like the time taken for the crossover is not same for all values of $\sigma$. Qualitatively, we can make a remark that with increasing $\sigma$, crossover time decreases. Our intuition tells us that the crossover time also should depend on the choice of the final temperature to which the system is quenched. Quantification of both will need even more extensive simulations. 

%Interestingly, our conclusion about the rapid growth during early times is in agreement with a study \cite{Corberi2019} in dimension $d=1$. Depending upon the value of $\sigma$, the early scaling period can extend over decades. It is possible that because of this reason the authors of the latter work missed the expected late time behavior with $\alpha=1/(2+\sigma)$. Nevertheless, we do not rule out the possibility of interesting dimensionality dependence of $\alpha$, leading to the observation in Ref. \cite{Corberi2019}. 

%\bibliographystyle{revtex}
%\bibliography{library}
\bibliography{library}

\end{document}